\documentstyle[11pt,epsf]{article}
\input epsf
\begin{document}
\baselineskip 100pt
\renewcommand{\baselinestretch}{1.5}
\renewcommand{\arraystretch}{0.666666666}
\parindent=0pt
{\large
\parskip.2in
\newcommand{\be}{\begin{equation}}
\newcommand{\ee}{\end{equation}}
\newcommand{\br}{\bar}
\newcommand{\fr}{\frac}
\newcommand{\lm}{\lambda}
\newcommand{\ra}{\rightarrow}
\newcommand{\al}{\alpha}
\newcommand{\bt}{\beta}
\newcommand{\pr}{\partial}
\newcommand{\hs}{\hspace{5mm}}
\newcommand{\up}{\upsilon}
\newcommand{\dg}{\dagger}
\newcommand{\ve}{\varepsilon}
\newcommand{\acc}{\\[3mm]}
\noindent

\hfill DTP\,96/09

\vskip 1truein
\bigskip
\begin{center}
{\bf Soliton Solutions and Nontrivial Scattering in an \\\
Integrable Chiral Model in (2+1) Dimensions.}
\end{center}

\vskip 1truein
\bigskip
\begin{center}
T. I{\small OANNIDOU}\\
{\sl Dept of Mathematical Sciences, University of Durham,\\
Durham DH1 3LE, UK}
\end{center}

\vskip 1truein
{\bf Abstract.}
The behaviour of solitons in integrable theories is  strongly constrained
by the integrability of the theory; i.e. by the  existence of an infinite
number of conserved quantities which these  theories are known to
possess.
One usually expects the scattering of solitons  in such
theories to be rather simple, i.e. trivial.
By contrast, in this paper we generate new  soliton
solutions for the planar integrable chiral model whose
scattering properties are highly nontrivial; more precisely, in head-on
collisions of $N$ indistinguishable solitons the scattering angle
(of the emerging structures relative to the incoming ones) is
$\pi/N$. \\
We also generate soliton-antisoliton solutions with elastic scattering;
in particular, a head-on collision of a soliton and an antisoliton
resulting in $90^0$ scattering.
\newpage

\noindent
{\bf {\rm {\bf I}}. Introduction.}

This paper studies certain exact soliton solutions of an
integrable system.
Before any detailed discussion, and to avoid confusion later on, it is
worthwhile clearing up a small point of terminology: the word {\it
solitons} was introduced by mathematicians to describe lumps of energy
which were stable to perturbations and did not change either velocity or
shape when colliding with each other.
However, in recent literature all sorts of localized energy
configurations have been called solitons.
We shall go along this looser definition.
By a soliton we shall mean a lump of energy that moves but we shall not
imply stability of the shape or the velocity or a simple behaviour in
collision.

An interesting problem is to look at the scattering properties of
two or more solitons colliding.
In some known systems with nontrivial
topology, the collision of two solitons is inelastic (some
radiation is emitted) and nontrivial (a head-on collision results in
$90^0$ scattering); all this has been observed
analytically \cite{R,S} and numerically \cite{Z}-\cite{MMR}.
One can construct explicit time-dependent solutions only in
very special, so-called integrable models.
Usually in these models extended objects
interact trivially, in the sense that they pass through each other with
no lasting change in velocity or shape (i.e., they behave as genuine
solitons).
Some examples in (2+1) dimensions are  the Kadomtsev-Petviashvili
equation \cite{NMPZ} and the integrable chiral model \cite{W}.
The last system is the subject of this paper and will be described below.
Until now, nontrivial scattering of solitons occurs mostly in
nonintegrable systems which is far from simple.
The question that arises is whether this type of
scattering can occur in  integrable models too.
There are some limited examples of integrable systems where soliton
dynamics can be nontrivial.
In (1+1) dimensions there many models which possess nontrivial
soliton-like solutions (cf. \cite{Mat}); like the boomeron solutions
\cite{CD}, which are solitons with time dependent velocities.
In (2+1) dimensions there are the dromion solutions \cite{BLMP} of the
Davey-Stewartson equation, which decay exponentially in both spatial
coordinates and interact in a nontrivial manner \cite{BS}; and the
soliton solutions \cite{GPS} of the Kadomtsev-Petviashvili equations,
whose scattering properties are highly nontrivial.

In the present work we are going to construct families of  soliton
solutions for the integrable (2+1)-dimensional chiral model and observe the
occurrence of different types of  behaviour.
This happens since the solitons in this system have internal degrees
of freedom which determine their  orientation in space;
do not affect the initial energy density; and are important in understanding
the evolution as a whole.
Therefore, they can interact either trivially or  nontrivially,
depending on the orientation of these internal parameters and on
the values of the impact parameter defined as the distance of
closest of approach between their centres in the absence of
interaction.
Namely, if two initial soliton-like structures are sent
towards each other at zero impact parameter, then, as most numerical
simulations have shown, the outgoing structures emerge at $90^0$.

To proceed further let us specify the system.
The modified SU(2) chiral model studied by Ward \cite{W} is given by the
field equation
\be
\pr^{\mu}(J^{-1}J_\mu)-\fr{1}{2}V_\al \varepsilon^{\al \mu
\nu}[J^{-1}J_\mu,J^{-1}J_\nu]=0.
\label{1}
\ee
Here $J$ takes values in SU(2) group and is thought of as a 2$\times$2
unitary matrix of functions of the spacetime coordinates on ${\bf R^{2+1}}$:
$x^\mu=(x^0,x^1,x^2)=(t, x, y)$ with $\mbox{det}\,J=1$.
Greek letters are spacetime indices, taking values 0, 1, 2, $\pr_\mu$
denotes partial differentiation with respect to $x^{\mu}$, while $J_\mu
\equiv \pr_\mu J$.
The quantity $\varepsilon^{\al \mu \nu}$ is the alternating tensor of
three indices with $\varepsilon^{012}=1$.
Finally, $V_\al$ is a unit vector in spacetime.
The conformal properties  of $V_\al$ determine whether
the symmetry group is SO(2) or SO(1,1) (depending on whether $V_\al$ is
timelike or spacelike).

Ward \cite{W} chooses $V_\al$ to have the components
$V_\al=(0,1,0)$, the spacelike case, so that (\ref{1}) is a chiral
equation with torsion term and has the same conserved energy-momentum
vector as the chiral field equation.
In fact, the corresponding energy density is
\be
{\cal E}=-\fr{1}{2}\mbox{tr}\left[(J^{-1}J_t)^2+(J^{-1}J_x)^2
+(J^{-1}J_y)^2\right].
\label{2}
\ee
Here tr denotes the matrix trace.
It should be emphasized that ${\cal E}$ is a positive-defined functional
of $J$, and hence a conserved energy exists which is the
integral of the energy density over the spacelike plane $x^0=\mbox{const}$.
 The boundary conditions are chosen so that the field configuration has
finite energy.
Consequently, we require that $J$ be everywhere smooth and that
\be
J=J_0+J_1(\theta)r^{-1}+O(r^{-2}),
\label{3}
\ee
at spatial infinity, with $x+iy=r\,e^{i\theta}$.
Here $J_0$ is a constant matrix, and $J_1$ depends only on $\theta$ (no
time dependence).

The ensuing system when $V_a$ is $i$ times a timelike vector instead of
spacelike has been studied in \cite{MZ}.
Equation (\ref{1}) admits solitons, localized in two dimensions,
with trivial scattering, i.e. each soliton suffers no change in velocity
and no phase shift upon scattering \cite{W,MZ}.
It is the purpose of this paper to construct new soliton solutions for
(\ref{1}), and investigate their scattering behaviour.
Such solutions are localized along the direction of motion; they are not
however, of constant size: their height, which corresponds to the
maximum of the energy density ${\cal E}$, is time dependent.

The rest of the paper is arranged as follows.
In the next section we shall briefly discuss the integrability properties
of (\ref{1}), and write down a family of multisoliton
solutions as configurations that are the limiting cases of the
ones already obtained using the standard method of {\it Riemann problem with
zeros} \cite{W}.
In section 3 we construct two families of multisoliton solutions with
nontrivial scattering; in particular, for the first one we prove that
in all head-on collisions the $N$ moving structures undergo
$\pi/N$ scattering.
In section 4 we construct a mixture of soliton-antisoliton solutions,
and in section 5 we discuss their dynamics and scattering properties.
We finish the paper with a short section containing our conclusions.

\noindent
{\bf {\rm {\bf II}}. Construction of Soliton Solutions.}

The integrable nature of equation (\ref{1}) means that there is a
variety of methods for constructing exact solutions.
Together with  {\it Riemann problem with zeros} \cite{W}, both {\it
twistor techniques} \cite{W2} and a full {\it inverse scattering
formalism} \cite{V} have been applied to the model.
This section indicates a general method for constructing soliton
solutions of the integrable chiral model (\ref{1}).
The technique is a variation of that in \cite{W,W1}, following a
pioneering idea of Zakharov and his collaborators \cite{BZ}.

The nonlinear equation (\ref{1}) is integrable in a sense that it may be
written  as the compatibility condition for the following linear
system
\be
\begin{array}{llcl}
L\psi\equiv(\lm\pr_x-\pr_u)\psi=A\psi,\\
M\psi\equiv(\lm\pr_\up-\pr_x)\psi=B\psi.
\label{4}
\end{array}
\ee
Here $\lm \in {\bf C}$, $(u,\up,x)$ are coordinates on ${\bf R^{2+1}}$
with $u=(t+y)/2$, $\up=(t-y)/2$, $A$ and $B$ are 2$\times$2
anti-hermitean trace-free matrices depending only on $(u,\up,x)$, and
$\psi(\lm,u,\up,x)$ is an unimodular $2\times2$ matrix function
satisfying the reality condition
\be
\psi(\lm,u,\up,x)\,\psi(\br{\lm},u,\up,x)^\dagger= I,
\label{5}
\ee
where bar denotes complex conjugate, $^\dagger$ denotes the complex
conjugate transpose matrix  and $I$ is the $2\times2$ identity  matrix.
The system (\ref{4}) is overdetermined, and in order
for a solution $\psi$ to exist, $A$ and $B$ have to satisfy the integrability
conditions, which are
\be
\begin{array}{c}
B_x=A_\up,\hs \hs A_x-B_u-[A,B]=0.
\label{6}
\end{array}
\ee
If we put $J(u,\up,x)=\psi(\lm=0,u,\up,x)^{-1}$ where $\psi$ is a
solution of the system (\ref{4}), we get by comparing (\ref{4}) and
(\ref{6}) that
\be
\begin{array}{c}
A=J^{-1}J_\up,\hs \hs B=J^{-1}J_x.
\end{array}
\ee
Therefore, the integrability condition for (\ref{4}) implies that there
exists a field $J$ which satisfies the equation of motion (\ref{1});
and moreover, the reality condition on $\psi$ ensures that $J$ is unitary.

Using the standard method of {\it Riemann problem with zeros},
in order to construct multi-soliton solution one may assume that the
function $\psi$ has simple poles in $\lm$, or in other words must
possess the form
\be
\psi(\lm)= I+\sum^{n}_{k=1}\fr{M_k}{\lm-\mu_k},
\label{8}
\ee
where $M_k$ are 2$\times$2 matrices independent of the complex parameter
$\lm$, $n$ is the number of solitons, and the complex parameter
$\mu_k$ determines the velocity of the $k$-th soliton.
The components of the matrix $M_k$ are given in terms of a
rational function  $f_k$ of the complex variable $\omega_k=x+\mu_k
u+\mu_k^{-1} \upsilon$.
(Roughly speaking, $f_k(\omega_k)$ describes the shape of the $k$-th
soliton). In fact, the matrix $M_k$ (cf. {\cite{W}) has the form
\be
M_k=-\sum^n_{l=1}(\Gamma^{-1})^{kl}\br{m}^l_a m^k_b,
\ee
with $\Gamma^{-1}$ the inverse of
\be
\Gamma^{kl}=\sum^2_{a=1}\left(\br{\mu}_k-\mu_l\right)^{-1}\br{m}_a^k
m^l_a.
\ee
Here $m_a^k$ are holomorphic functions of $\omega_k$, given by
$m_a^k=(m_1^k, m_2^k)=(1,f_k)$.
These solitons pass each other without any change of direction or phase
shift.
Infinite energy  extended wave solutions \cite{L1} may be constructed by
taking $f_k$ to be an exponential function of $\omega_k$.
Such extended wave solutions suffer a phase shift upon scattering,
although again there  is no change in velocity.

All this assumes that the parameters $\mu_k$ are distinct, and also
$\bar{\mu}_k \neq \mu_l$ for all $k$,  $l$.
In this paper examples are given of two generalizations of these
constructions: one involving higher-order poles in $\mu_k$, and the other
where $\bar{\mu}_k \neq \mu_l$.

Let us look at an example in which the function $\psi$ has a a double
pole in $\lm$, and no other poles.
So we take $\psi$ to have the form
\be
\psi=I+\sum^{2}_{k=1}\fr{R_k}{(\lm-\mu)^k},
\label{11}
\ee
where $R_k$ are 2$\times$2 matrices independent of $\lm$.
[This hypothesis can be generalized by taking the function $\psi$ to
have  a pole of order $n$ in $\lm$.]

It has been proved \cite{W1} that $\psi$ given by (\ref{11})
satisfies the reality condition (\ref{5}) if and only if it factorizes as
\be
\psi(\lm)=\left(I-\fr{(\br{\mu}-\mu)}{(\lm-\mu)}\fr{q_1^\dg
\otimes q_1}{\|q_1\|^2}\right)\left(I-\fr{(\br{\mu}-\mu)}
{(\lm-\mu)}\fr{q_2^\dg\otimes q_2}{\|q_2\|^2}\right),
\label{10}
\ee
where  $q_k$ are two-dimensional row vectors and $\|q_k\|^2=q_k
\cdot q_k^\dg$.

The $q_k$ have to satisfy a condition, which amounts to saying the
matrices $A=(L\psi)\psi^{-1}$ and $B=(M\psi)\psi^{-1}$ are independent of
$\lm$.
One way of obtaining $q_k$ with this property is as a limit of the
simple-pole case (\ref{8}) with $n=2$.
The idea is to take a limit $\mu_k \ra \mu$.
In order to end up with a smooth solution $\psi$ for all ($u, \up, x$),
it is necessary that $f_2(\omega_2)-f_1(\omega_1) \ra 0$ in this limit.

In our case, with $n=2$, we put $\mu_1=\mu+\ve$, $\mu_2=\mu-\ve$ and
write $f_1(\omega_1)=f(\omega_1)$, $f_2(\omega_2)=f(\omega_2)$, with $f$
being a rational function of one variable.
In the limit $\ve \ra 0$, $\psi$ has the form (\ref{10}), with
\be
\begin{array}{llcl}
q_1=(1+|f|^2)(1,f)+\varphi \, (\br{\mu}-\mu)(\br{f},-1),\\
q_2=(1,f).
\label{13}
\end{array}
\ee
Here $f$ is a rational function of  $\omega=x+\mu u+\mu^{-1}
\up$, $\varphi=(u-\mu^{-2}\up)\,f^{\prime}(\omega),$
while $f^\prime(\omega)$ denotes the derivative of $f(\omega)$ with
respect to its argument.
As a result, we have a solution $J=\psi(\lm=0)^{-1}$ depending on the
complex parameter $\mu$ and on the arbitrary function $f$.
In fact, it  has the form of the following product
\be
J=\left(I+\fr{(\br{\mu}-\mu)}{\mu}\fr{q_2^\dg \otimes
q_2}{\|q_2\|^2}\right)\left( I+\fr{(\br{\mu}-\mu)}{\mu}\fr{q_1^\dg
\otimes q_1}{\|q_1\|^2}\right),
\label{16}
\ee
with $q_k$ given by (\ref{13}).
Notice that $J$ takes values in SU(2); is smooth everywhere on ${\bf
R^{2+1}}$ (mainly because, the two vectors $q_1$ and $q_2$ are nowhere
zero); it satisfies the boundary condition (\ref{3}); and the equation of
motion (\ref{1}).

To start with, and in order to illustrate the above family of soliton
solutions, let us examine two simple cases, by giving specific values
to the parameters $\mu$ and  $f(\omega)$.
[The complex parameter $\mu$ determines the velocity of the
``centre-of-mass" of the system.]

\begin{itemize}
\item Let us take $\mu=i$ (which corresponds to the ``centre-of-mass"
of the system being stationary) and $f(\omega)=\omega$, thus $\omega=z$
and $\varphi=t$, where $z=x+i y$; $r^2=z\br{z}$.
Therefore the row vectors (\ref{13}), become
\be
\begin{array}{llcl}
q_1=(1+r^2)(1,z)-2it(\br{z},-1),\\
q_2=(1,z).
\end{array}
\ee
In this time-dependent solution, for $t$ negative, a ring structure with
reducing radius is obtained, which deforms to a single peak at $t=0$ and
thereafter expands again to a ring.
Figure 1 presents few pictures of the corresponding energy density  at some
representative values of time.
Ring structures  occur in the soliton  scattering of many nonintegrable
planar systems \cite{Z,SR} and are an approximation of two solitons.

This picture can be confirmed by looking at the energy density of the
solution, which is
\be
{\cal E}=16\fr{r^4+2r^2+4t^2(2r^2+1)+1}{[r^4+2r^2+4t^2+1]^2}.
\label{18}
\ee
Notice that the energy density is time-reversible and
rotationally symmetric (see below).
For large (positive) $t$, the height of the ring (maximum of ${\cal
E}$) is proportional to $1/t$, while its radius is
proportional to $\sqrt{t}$.

\item Accordingly, let us take $\mu=i$ and $f(\omega)=\omega^2$.
Thus, the row vectors (\ref{13}) are
\be
\begin{array}{llcl}
q_1=(1+r^4)(1,z^2)-4itz(\br{z}^2,-1),\\
q_2=(1,z^2).
\end{array}
\ee
Here, for negative $t$, a single peak
occurs with an additional ring, which changes to a ring structure at
$t=0$ and reverts back to the original form, for positive $t$ (see
Figure 2).
However, these rings are not radiation since they travel with speed
less than that of light.
In fact, for large (positive) $t$, their velocity  is approximately
proportional to $t^{-2/3}$.
[Note that we have set the velocity  of the light,  $c$, equal to the
unity, so that in all our  calculations  we can use  dimensionless
quantities.]

This leads to an energy density, which is
\be
{\cal E}=64\fr{r^{10}+18t^2r^8+2r^6+4t^2r^4+r^2+2t^2}{\left
[r^8+2r^4+16t^2r^2+1\right]^2}.
\label{20}
\ee
Again, ${\cal E}$ has the same symmetries as in (\ref{18}).
For large (positive) $t$, the height of the soliton peak is proportional
to $t^2$ and its radius is proportional to $1/t$; while the soliton  ring
spread out, becoming broader and broader, with  height proportional  to
$t^{-2/3}$ and radius proportional to $t^{1/3}$.
\end{itemize}

Finally, a general concluding remark should be made.
Although (\ref{1}) is not rotationally symmetric in the
$x y$-plane; when  $f(z)=z^p$ the field $J$ (\ref{13},\ref{16}) is
invariant under the transformation $z \ra e^{i \phi} z$, since
\be
J \ra J^{\prime}=\left( \begin{array}{llcl} e^{i \phi p/2} & 0\\ 0 &  e^{-i
\phi p/2} \end{array} \right) J \left( \begin{array}{llcl} e^{-i \phi
p/2} & 0\\ 0 &  e^{i \phi p/2} \end{array} \right).
\ee
This transformation does not affect the equation of motion (\ref{1}) due
to the chiral symmetry $J \ra {\cal \kappa}J{\cal \tau}$ where ${\cal
\kappa}$ and ${\cal \tau}$ are constant SU(2) matrices.
The main features of this time-dependent solution may be inferred as follow.
If $r$ is large, the field $J$ is close to its asymptotic
value $J_0$, as long as $2tf^{\prime}/|f|^2 \ra 0$.
But as $2t|f^{\prime}|/|f|^2 \approx 1$, $J$ departs from its asymptotic
value $J_0$ and a ring structure emerge with radius proportional to
$\left(2tp\right)^{1/(p+1)}$.

\noindent
{\bf {\rm {\bf III}}. Soliton-Soliton Scattering.}

We now move on to the more interesting question of scattering processes.
In fact, we will use the method of section 2 to construct
solutions of (\ref{1}) representing scattering solitons.
We will see that, in all head-on collisions of $N$ moving solitons the
scattering angle is $\pi/N$.
Moreover, when the $N$  solitons are very close together, and in
particular, when they are on top of each other the $N$ lumps  which
represent them merge together to form a ring-like structure.
Then, instead of moving towards the centre, they emerge from the ring in
a direction that bisects the angle formed by the incoming ones.
As we have already  mentioned this nontrivial  scattering is not usual
in an integrable theory, but is exceptional.

The scattering solutions arise if we take a solution of the simple-pole
case (\ref{8}) with $n=2$, put $\mu_1=\mu+\ve$, $\mu_2=\mu-\ve$ and take
the limit $\ve \ra 0$.
The constraint $f_2(\omega_2)-f_1(\omega_1) \ra 0$ as $\ve \ra 0$ has to be
imposed, in order for the resulting solution $\psi$ to be smooth for all
$(u,\up,x)$.
So let us write $f_1(\omega_1)=f(\omega_1)+\ve h(\omega_1)$,
$f_2(\omega_2)=f(\omega_2)-\ve h(\omega_2)$, where $f$ and
$h$ are both rational functions of one variable (the examples of the
previous section had $h=0$).
Once again $J$ is given by (\ref{16}), with the two-vectors
$q_k$ given by
\be
\begin{array}{llcl}
q_1=(1+|f|^2)(1,f)+\vartheta\, (\br{\mu}-\mu)(\br{f},-1),\\
q_2=(1,f),
\end{array}
\label{22}
\ee
where $\vartheta=\varphi+h(\omega)$.
So this solution belongs to a large family, since one may take $f$
and $h$ to be any  rational meromorphic functions of $\omega$.
Note that $J$ is smooth on ${\bf R^{2+1}}$ and satisfies its boundary
condition, irrespective of the choice of $f$ and $h$.

It may seem strange that one can take the limit of
a family of soliton solutions with trivial scattering, and
obtain a new one with nontrivial scattering.
Thus, it is interesting to study how the solitons are affected by
varying $\ve$.
To do so, let us take a solution of the simple-pole case (\ref{8}) with
$n=2$, put $\mu_1=i+ \ve$, $\mu_2=i- \ve$, while taking $f_k=\omega_k$;
and study how the configuration of the two initial well separated solitons
changes as $\ve \ra 0$ at a fixed time ($t=-15$).
Figure 3 shows that as $\ve \ra 0$ the solitons disperse, shift and
interact with each other. In other words, their internal degrees of
freedom as well as the impact parameter change in this limit, making
the process highly nontrivial.

As an example, let us present two typical  cases.

\begin{itemize}
\item Let us take $\mu=i$, $f(\omega)=\omega$
and $h(\omega)=\omega^3$; thus $\vartheta=t+z^3$.
For  $r$ large,  $J$  is equal to  its
asymptotic value $J_0$, as long as $\vartheta/z^3=1+t/z^3\approx 1$, but as
$z$ approaches any of the three cube roots of $-t$, then
$\vartheta \ra 0$, while $J$ departs from $J_0$, and
three localized solitons emerge. For $t$ negative, the three solitons are
approximately at the points:
$\left((-t)^{1/3},0\right)$,  $\left(-(-t)^{1/3},\pm
\sqrt{3}\,(-t)^{1/3}\right)$; while for $t$ positive,
the solitons are at $\left(-t^{1/3},0\right)$,
$\left(t^{1/3},\pm \sqrt{3}\,t^{1/3}\right)$.

More information can be deduced from the energy density, which is
\be
\begin{array}{llcl}
{\cal E} =&16\,\big[2r^8+16r^6+19r^4+2r^2(1+8x
y^2t)+4t^2(1+2r^2)+1+8x y^4t\\
 &  -8x^5t-16tx(x^2-y^2)\big]/\big[4r^6+r^4+2r^2+4t^2+1+8tx(x^2-3y^2)
\big]^2.\\
\end{array}
\ee
The density ${\cal E}$ is symmetric under the interchange $t \mapsto
-t$, $x \mapsto -x$ and $y \mapsto -y$.
For small (negative)  $t$, the solitons form an intermediate state
having the shape of a ring with three maxima on the direction of the
incoming solitons which deforms to a circularly-symmetric
ring at $t=0$ and then energy seems to flow
around, until three other maxima are formed in the transverse direction,
for small (positive) $t$.
Figure 4 shows clearly  the intermediate states with three  maxima.
The three new  maxima then  give rise to three new solitons emerging at
$60^0$ to the original direction of motion.
During the intermediate phase solitons lose their identity.

Finally something has to be said about their size.
For large (positive) $t$, their height is proportional to
$t^{-4/3}$, their radius is proportional to $t^{1/3}$, while their speed is
proportional to $t^{-2/3}$: therefore, they spread out and slow down.

\item Accordingly, let us take  $\mu=i$ while
choose $f(\omega)=\omega^2$ and $h(\omega)=\omega^3$.
Here $J$ departs from its asymptotic value $J_0$, when
$z$ approaches the  values  $\pm \sqrt{-2t}$ or
 zero (since $\vartheta=z(2t+z^2)\ra 0$);
and (again) three localized solitons emerge.
In this case though, if $t$ is negative, all three of them are on the
$x$-axis at $x \approx \pm \sqrt{-2t}$ and at the origin; while if $t$
is positive, they are on the  $y$-axis
at $y \approx \pm \sqrt{2t}$ and at the origin.
So the picture consists of three solitons: a static one at the origin,
with the other two accelerating towards the origin, scattering at
right angles and then decelerating as they separate.

This can be observed from the energy density, which is
\be
\begin{array}{llcl}
{\cal E}
= &32\big[r^{12}+2r^2(r^8+r^6+1)+36t^2r^8+4r^6+9r^4+8t^2r^4+4t^2
+12t(x^{10}-y^{10}) \\
 &+4t(x^2-y^2)(3+2x^2y^2+6x^4y^4)+4t(x^6-y^6)(9x^2y^2-2)-y^{10})\big]/\big
[r^8+4r^6\\
 &+2r^4+16tr^2(t+x^2-y^2)+1\big]^2. \\
\end{array}
\ee
Here ${\cal E}$ is symmetric under the interchange $t \mapsto
-t$, $x \rightleftharpoons y$; therefore the collision is time
symmetric, with the only effect the $90^0$ scattering (no phase shift; no
radiation).
For large (positive) $t$, the height of the static soliton is proportional
to $t^2$ and its radius is  proportional to $1/t$; while the moving solitons
expand with height proportional to $t^{-2/3}$ and  radius proportional to
$t^{1/3}$.

In Figure 5 we present some pictures of the total energy densities of
three solitons during a typical nontrivial evolution.
\end{itemize}
In principle one should be able to visualize the emerging soliton
structures when $f(\omega)=\omega^p$ and $h(\omega)=\omega^q$, i.e.
are rational of degree $p,q \in {\bf N}$, respectively.
In fact, for $q>p$ the configuration consists of
$(p-1)$  static solitons at the ``centre-of-mass" of the system (if
more than one, a ring structure is formed) accompanied by $N=q-p+1$
solitons  accelerating  towards the ones in the middle, scattering at an
angle of $\pi/N$, and then decelerating as they separate.
This follows from the fact that the field $J$ departs from its asymptotic
value $J_0$ when
$\vartheta=\omega^{(p-1)}(p\,(u-\mu^{-2}\up)+\omega^N) \ra 0$, which
is true when either $\omega^{(p-1)}=0$ or
$\omega^N+p\,(u-\mu^{-2}\up)=0$; and this is
approximately where the solitons are located.

We conclude this section by investigating the corresponding case
where $\psi(\lm)$ has a triple pole (and no others). Therefore, it is
taken to have the form
\be
\psi(\lm)=I+\sum^{3}_{k=1}\fr{{\cal R}_2}{(\lm-\mu)^k}.
\ee
As we have already mentioned, the reality condition (\ref{5}) is
satisfied if and only if $\psi$ factorizes into three simple factors of
the following type
\be
\psi(\lm)=i\,\left(I-\fr{(\bar{\mu}-\mu)}{(\lm-\mu)}\fr{q_1^\dg
\otimes q_1}{\|q_1\|^2}\right)\left(I-\fr{(\br{\mu}-\mu)}
{(\lm-\mu)}\fr{q_2^\dg\otimes q_2}{\|q_2\|^2}\right)
\left(I-\fr{(\bar{\mu}-\mu)}{(\lm-\mu)}\fr{q_3^\dg
\otimes q_3}{\|q_3\|^2}\right),
\label{tr}
\ee
for some two-vectors $q_k$.
The requirement that the matrices $A=(L\psi)\psi^{-1}$ and
$B=(M\psi)\psi^{-1}$ should be independent of $\lm$ imposes differential
equations on $q_k$; which are three nonlinear equations, and it seems
difficult to find their general solution.

One way of proceeding is to take a solution for the simple-pole case
(\ref{8}) with $n=3$, put $\mu_1=i+\ve$, $\mu_2=i$, $\mu_3=i-\ve$ and take
the limit $\ve \ra 0$.
In order to obtain a smooth solution $\psi$ for all ($u,\up,x$), it is
necessary that $f_1(\omega_1)-f_2(\omega_2) \ra 0$,
$f_1(\omega_1)-f_3(\omega_3) \ra 0$, $f_2(\omega_2)-f_3(\omega_3) \ra 0$
as $\ve \ra 0$.
So let us write $f_1(\omega_1)=f(\omega_1)+\ve\, h(\omega_1)+\ve^2\,
g(\omega_1)$,  $f_2(\omega_2)=f(\omega_2)$,
$f_3(\omega_3)=f(\omega_3)-\ve\, h(\omega_3)+\ve^2\, g(\omega_3)$,
where $f$, $h$ and $g$ are rational functions of one variable.
On taking the limit, we obtain a $\psi$ of the form (\ref{tr}), smooth on
${\bf R^{2+1}}$ and such that the matrices $A$ and $B$ be independent of
$\lm$.

Consequently, $J=\psi(0)^{-1}$ is a smooth solution of (\ref{1}) of the form
\be
J=i\left(I-\fr{2q_3^\dg
\otimes q_3}{\|q_3\|^2}\right)\left(I-\fr{2q_2^\dg
\otimes q_2}{\|q_2\|^2}\right)\left( I-\fr{2q_1^\dg
\otimes q_1}{\|q_1\|^2}\right),
\label{sol}
\ee
with $q_k$ being in terms of $f(z)$, $h(z)$ and $g(z)$ by
\be
\begin{array}{llcl}
q_1=(1+|f|^2)^2(1,f)-4i(b+id)(1+|f|^2)(\br{f},-1)-4b^2(\br{f}^2,-\br{f}
-2i\br{b})-8id\br{b}(1,f),\\
q_2=(1+|f|^2)(1,f)-2ib\,(\br{f},-1),\\
q_3=(1,f),
\end{array}
\ee
where $b=t f^\prime(z)+h(z)$ and $d=t^2f^{\prime
\prime}(z)/2+i(t-y)f^\prime(z)/2+th^\prime(z)+g(z)$.
Note that the two-vectors $q_2$, $q_3$ here correspond to the
ones given by (\ref{22}) for $\mu=i$, respectively.

Let us examine a sample example of this solution, since we may
take $f$, $h$ and $g$ to be any rational meromorphic function of $z$.
\begin{itemize}
\item Let us take $f(z)=0$, $h(z)=z$
and $g(z)=z^2$; thus $b=z$ and $d=t+z^2$.
This solution consists of two solitons coming in along the $y$-axis
merging to form a peak at the origin and then two new solitons emerging
along the $x$-axis.
Figure 6 illustrates what happens near $t=0$.

The energy density of the system is,
\be
{\cal E}=32\fr{80r^4+32(r^2+t^2)+256t^2r^2-64t(x^2-y^2)+128tyr^2-8y+3}
{[32r^4+12r^2-16yr^2+16t^2+16ty+32t(x^2-y^2)+1]^2},
\ee
which has a reflection symmetry around the  $x$-axis.
For large (positive) $t$, ${\cal E}$ is peaked at two points on the
$y$-axis, namely $y\approx \pm \sqrt{t}$. Moreover, the height of the
corresponding solitons is proportional to $1/t$, and their radius is
proportional to $\sqrt{t}$; which means that the y-axis asymmetry
vanishes at $t \ra \infty$.
\end{itemize}

\noindent
{\bf {\rm {\bf IV}}. Construction of Soliton-Antisoliton Solutions.}

In this section we construct a large family of solutions
which as we will argue later, can be though of as representing
soliton-antisoliton field configurations.
Roughly speaking, solitons correspond to  $f$ being a function of
the variable $z$, and antisolitons correspond to a function of $\br{z}$.

One way to generate a soliton-antisoliton solution of (\ref{1}),
is to assume that $\psi(\lm)$ has
the form
\be
\psi(\lm)={\it I}+\fr{n^1\otimes m^1} {(\lm-i)}+\fr{n^2\otimes m^2}
{(\lm+i)}.
\label{24}
\ee
Here  $n^k$, $m^k$ for $k=1,2$ are complex-valued two-vector functions of
$(t,z,\br{z})$ (not depending on $\lm$).

The idea is to find the $n^1_1,...,m_1^1,...$ such that the reality
condition (\ref{5}) holds, and such that the matrices
$A=(L\psi)\psi^{-1}$ and $B=(M\psi)\psi^{-1}$ are independent of $\lm$.
One way of proceeding is to take the solution (\ref{8})
with $n=2$, put  $\mu_1=i+\varepsilon$, $\mu_2=-i-\varepsilon$ and
take the limit $\varepsilon \ra 0$.
In order for the resulting $\psi$ to be smooth on ${\bf R^{2+1}}$ it is
necessary to take $f_1=f(\omega_1)$,
$f_2=-1/\br{f}(\omega_2)-\varepsilon\, h(\omega_2)$, where $f$ and $h$
are rational functions of one variable.
On taking the limit $\varepsilon \ra 0$, we then obtain a $\psi$ as in
(\ref{24}) with $m^k=(m^k_1,m^k_2)$ being holomorphic functions of $z$
(or $\br{z}$), through the relations $m^1=(1,f)$, $m^2=(-\br{f},1)$, while
\begin{eqnarray}
n^1&=&\fr{2i(1+|f|^2)}{(1+|f|^2)^2+|w|^2}\,\br{m}^1+\fr{2\br{w}}
{(1+|f|^2)^2+|w|^2}\,\br{m}^2,\nonumber \\
n^2&=&-\fr{2w}{(1+|f|^2)^2+|w|^2}\,\br{m}^1-\fr{2i(1+|f|^2)}
{(1+|f|^2)^2+|w|^2}\,\br{m}^2,
\end{eqnarray}
with
\be
w\equiv \br{h}f^2+2t f^\prime.
\label{w}
\ee
So we generate a solution $J=\psi(\lm=0)^{-1}$,
which depends on the two arbitrary rational functions $f=f(z)$ and
$h=h(\br{z})$.
This solution has the form
\be
J=\fr{1}{(1+|f|^2)^2+|w|^2}\left[\begin{array}{llcl}
|w|^2+2i(f\br{w}+\br{f}w)-(1+|f|^2)^2
\hs \hs  -2i\left(w-f^2\br{w}\right)\acc
-2i\left(\br{w}-\br{f}^2w\right) \hs \hs
|w|^2-2i(f\br{w}+\br{f}w)-(1+|f|^2)^2
\end{array}\right],\\
\label{27}
\ee
with $w$ given by (\ref{w}).
In general, by taking $f(z)=z^p$ and $h(\br{z})=\br{z}^q$ where $p$
is a positive integer and $q$ is a non-negative integer; the
energy, obtained by integrating (\ref{2}), is $E=(2p+q)8\pi$.
Roughly speaking, the solution looks like $(2p+q)$ lumps  at arbitrary
positions in the $x y$-plane; which as we are going to see are a
combination of solitons and antisolitons.

A topological charge may be defined for the field $J$ (\ref{27})
by exploiting the connection of it with the $\rm{O}(3)$ $\sigma$-model.
The unmodified chiral model [i.e., (1) with $V^{\al}=(0,0,0)$] is
equivalent to the $\rm{O}(4)$ $\sigma$-model \cite{S1} through the relation
\be
J=I\, \phi_0+\mbox{\boldmath $i \sigma \cdot \phi$},
\label{28}
\ee
where \(\mbox{\boldmath $\sigma$}\) are the usual Pauli matrices, and
$(\phi_0, \mbox{\boldmath $\phi$} )=(\phi_0,\phi_1,\phi_2,\phi_3)$ is a
four vector of real fields that are constrained to lie on ${\bf
S^3}$, i.e. \(\phi_0^2+\mbox{\boldmath $\phi \cdot \phi$}=1\).
The only static finite energy solutions of  the $\rm{O}(4)$ $\sigma$-model
correspond to the embedding of the $\rm{O}(3)$ $\sigma$-model \cite{BG}.
Therefore the only static solutions of (1) are the $\rm{O}(3)$
embeddings that we shall describe.
This is because for the one-soliton solution (static or Lorentz boosted
in the $y$-axis) the term in
(\ref{1}) proportional to $V^\al$ is zero, so the system  behaves like the
$\rm{O}(4)$ model, for which the $\rm{O}(3)$ embedding is totally geodesic.
[However, for time-dependent configurations, the term proportional to
$V^\al$ is non-zero and will affect the evolution of the field, which
will in general not lie in an  $\rm{O}(3)$ subspace of $\rm{O}(4)$.]

To proceed further, let us mention the topological aspects
of the $\rm{O}(3)$ and  $\rm{O}$(4) $\sigma$-models.
In studying soliton-like solutions, we require that the field
configuration has finite energy.
This implies that the field must take the same value at all points of
spatial infinity, so that space is compactified from ${\bf R^2}$ to ${\bf
S^2}$.
At fixed time, the field is a map from ${\bf S^2}$ into the target space.
Now for the $\rm{O}$(3) model, the field is a map \(\mbox{\boldmath $\phi$}:
{\bf S^2} \ra  {\bf S^2}$, and  due to the homotopy relation
\be
\pi_2 ({\bf S^2})={\bf Z},
\ee
such maps are  clasified by an integer winding
number ${\cal N}$ which is a conserved topological charge.
An expression for this charge is given by
\be
{\cal
N}=(8\pi)^{-1}\int\epsilon_{ij}\,\mbox{\boldmath$\phi$}\cdot(\pr_i
\mbox{\boldmath $\phi$}\wedge\pr_j \mbox{\boldmath $\phi$})\,d^2x,
\ee
where $i=1,2$ with $x^i=(x,y)$.

Although, for the $\rm{O}$(4) model [the same argument is valid for
(\ref{1}) due to the topological aspects of the theory] the
field at fixed time is a map $(\phi_0,\mbox{\boldmath $\phi$}): {\bf
S^2} \ra  {\bf S^3}$ and the corresponding homotopy relation is
\be
\pi_2 ( {\bf S^3})=0,
\ee
so there is no winding number.
However, for soliton solutions that correspond to some initial embedding
of $\rm{O}$(3) space into $\rm{O}$(4), there is a useful
topological  quantity, as we are going to see.

Consider the $\rm{O}$(4) configuration which at some time corresponds to an
$\rm{O}$(3) embedding, which we choose to be $\phi_0=0$ for definiteness.
At this time the field is restricted to an ${\bf S^2}$ equator of the
possible ${\bf S^3}$ target space.
Suppose that the field never maps to the anti-podal points
$\{{\cal A}_1,{\cal A}_2\}=\{\phi_0=1,\phi_0=-1\}$ at any time, so the
target space is ${\bf S^3_0}={\bf S^3}-\{{\cal A}_1,{\cal A}_2\}$.
Now  ${\bf S^3_0} \approx {\bf S^2} \times {\bf R}$, and thus we have the
homotopy relation
\be
\pi_2 ( {\bf S^3_0})=\pi_2 ( {\bf S^2} \times {\bf R})=\pi_2 ( {\bf S^2})
\oplus \pi_2 ({\bf R})={\bf Z},
\ee
and therefore a topological winding number exists.
An expression for this winding number is easy to give, since it is the
winding number of the map after projection onto the chosen ${\bf S^2}$
equator, i.e.
\be
{\cal N}^\prime=(8\pi)^{-1}\int\epsilon_{ij}\,\mbox{\boldmath
$\phi^\prime$}\cdot (\pr_i\mbox{\boldmath $\phi^\prime$}\wedge\pr_j
\mbox{\boldmath $\phi^\prime$})\,d^2x,
\ee
where $\mbox{\boldmath $\phi^\prime$}=\mbox{\boldmath
$\phi$}/|\mbox{\boldmath $\phi$}|$.
If the field does map to the anti-podal points $\{{\cal A}_1,{\cal A}_2\}$
at some time the winding number is ill defined at this time and if
considered as a function  of time ${\cal N}^\prime$ will be integer valued
but may suffer discontinuous jumps as the field moves through the
anti-podal points.
In the following examples, before comparing the solution $J$ given by
(\ref{27}) with the ${\rm O}$(3) embedding it is
convenient to perform the transformation $J \ra M\,J$ with
$M=(\sqrt{2})^{-1}\left[\begin{array}{llcl}
 \,\,\,\, 1 \hs 1\acc -1 \hs 1 \end{array}\right]$
so that the evolution of the field remains close to the $\rm{O}$(3)
embedding.

\noindent
{\bf {\rm {\bf V}}. Soliton-Antisoliton Scattering.}

Usually in the nonintegrable models, there is an attractive force between
solitons of opposite topological  charge.
In fact, if the solitons and antisolitons are well separated, then
they attract  each other and eventually annihilate into a wave of pure
radiation which spreads with the velocity of light \cite{Z,PSZ}.
However, the interaction forces between solitons and antisolitons
do depend on their configuration; in particular, they depend on the
relative orientation between them in the internal space.
Therefore,  the cross section for the soliton-antisoliton elastic
scattering is non-zero.
[In the real world, the proton-antiproton elastic scattering
is seen in a reasonable fraction of the cases.]
This is the first example for which there has been constructed an explicit
(since the system is integrable) solution of elastic
soliton-antisoliton  scattering in either integrable or nonintegrable
model.
As a  result, it provides a major  link between soliton dynamics
in integrable and nonintegrable systems.

The evolution is initially similar to the numerical results obtained through
the connection of the integrable chiral model (\ref{1}) with the
$\rm{O}$(3)  $\sigma$-model \cite{S1}.
In particular, a soliton and an antisoliton are moving along the $x$-axis
towards each other at an accelerating rate until they merge at the
origin and form a peak.
Note  that a peak is formed rather than a ring since the energy is
mainly kinetic when a soliton and an antisoliton merge.
However, rather than the peak dissipating in a wave of radiation
it now reforms into two new structures which undergo $90^0$ scattering.
In general, in all head-on collisions of $N$ moving soliton and
antisoliton objects, the scattering angle is $\pi/N$ degrees relative to the
initial direction of motion.

Next we looked at two cases corresponding to the mixtures of solitons and
antisolitons.
[The configurations given by (\ref{27}) when $h(\bar{z})=0$
are equivalent to the ones obtained from (\ref{13},\ref{16}) when
$f(z)=z^p$.]

\begin{itemize}
\item First, let us take $f(z)=z$ and $h(\br{z})=1$.
Roughly speaking, if $r$ is large, $J$ is close to its
asymptotic value  $J_0$, as long as $w/z^2=1+2t/z^2\approx 1$; but as $z$
approaches $\pm\sqrt{-2t}$ then $w\ra 0$, and $J$ departs from its asymptotic
value: this is where the two structures are located.
More precisely, for  negative $t$, the
two objects are on the $x$-axis, approximately at $x\approx \pm \sqrt{-2t}$;
while for positive $t$, they are on the $y$-axis, approximately at
$y\approx \pm  \sqrt{2t}$.
Figure 7 illustrates what happens near $t=0$.

The picture is consistent with the properties of the energy density of
the solution, which is
\be
{\cal E}=16\fr{2r^4+4r^2+4t^2(1+2r^2)-4t(x^2-y^2)+1}
{[2r^4+2r^2+4t(x^2-y^2)+4t^2+1]^2}.
\label{30}
\ee
Note the symmetry of ${\cal E}$ under the interchange $t
\mapsto-t$, $x \rightleftharpoons y$; the time symmetry of
the density confirms the lack of radiation.
The corresponding localized structures are not however of constant size:
for large (positive) $t$, their height  is proportional to $1/t$, while
their radius is proportional to $\sqrt{t}$.

The projected topological charge ${\cal N}^\prime$ is zero
throughout the scattering process; while the projected
topological density $q^{\prime}$, i.e.
\be
{\cal N}^\prime=\int q^\prime\, dx\,dy,
\ee
has an almost identical distribution (up to a scale) to
that of the energy density (see Figure 8(a)).
Therefore, the configuration represents  a soliton and an antisoliton
which are clearly visible as distinct structures having respectively +1
and -1 units of topological charge concentrated in a singe lump.
\end{itemize}

Equation (\ref{1}) is not Lorentz invariant and indeed is not even
radially symmetric due to the presence of the vector $V_a$ which picks
out a particular direction in space, and therefore one may expect to
find different scattering behaviour for more general solutions; e.g.,
when the soliton and  the antisoliton are moving along the $x$-axis
rather that the $y$-axis.
However, this is not true since (\ref{1}) is a
reduction of the self-dual Yang-Mills equation in ${\bf R^{2+2}}$ which
does have an SO(1,2) symmetry.
Therefore, the SO(2) symmetry of the Yang-Mills system means that any
given solution $J$, can in principle be converted  to  gauge fields by
performing a coordinate rotation (together with a gauge transformation)
and then recover the corresponding $J^{\prime}$ which will describe the
same  solution as $J$ but with a rotated coordinate system.
Indeed, this is what happens by taking
\be
\begin{array}{c}
f(z)=e^{(2i\phi)}z,\hs \hs h(\br{z})=1,
\end{array}
\ee
where $\phi$ is an angle in the $x y$-plane.
This picture presents a rotated version through any angle $\phi$ in the
$x y$-plane of the original one (i.e., Figure 7).

\begin{itemize}
\item Finally, let us take  $f(z)=z$ and $h(\br{z})=\br{z}$.
The corresponding configuration consists of one soliton
and two antisolitons (see Figure 8(b)).

It is interesting to look at the time dependence of various energies in
each process.
The total energy, of course, is constant and it is the spatial integral
of the following energy density
\be
\begin{array}{llcl}
{\cal E}&=&8\big[r^8+8r^6+11r^4+4r^2-8x^5t+16ty^2(x^3+t)+8t^2+48xy^2t+2\\
& &-16x^2t(x-t)+24xty^4\big]/\big[r^6+r^4+2r^2+4t^2+4tx^3-12x
y^2t+1\big]^2.\\
\end{array}
\ee
Obviously, the energy density ${\cal E}$ is symmetric under the
interchange $t \mapsto -t$, $x \mapsto -x$ and $y \mapsto -y$, only.
Again all three structures come together forming a bell-like structure
and then  emerge at an angle of $60^0$ with respect
to the original direction.
However, by looking at the maximum of ${\cal E}$ we
observe that,
for large (positive) $t$, the height of the localized structures is
proportional to $t^{-4/3}$, while their radius is proportional to $t^{1/3}$;
thus they spread out as they move apart.

Figure 9 shows the results of a head-on collision of the one-soliton
two-antisoliton system.
\end{itemize}

Let us conclude with the observation that, by taking $f(z)=z^p$ and
$h(\br{z})=\br{z}^q$, $J$ departs from its asymptotic value
$J_0$ when $w=z^{p-1}(2tp+z^N)\ra 0$ with $N=p+q+1$, which is true when
either $z^{(p-1)}=0$ or $2tp+z^N=0$: this is approximately where the
lumps are located.
Therefore, $J$ represents a family of soliton-antisoliton solution
which consists of ($p-1$) static soliton-like objects at the origin,
with $N$ others accelerating towards them, scattering at an angle of
$\pi/N$, and then decelerating as they separate.

\noindent
{\bf {\rm {\bf VI}}. Conclusion.}

The infinite number of conservation laws associated with a given
integrable system place severe constraints upon possible soliton dynamics.
The construction of exact analytic multisoliton solutions with trivial
scattering properties is a result of such integrability properties.
In this paper new soliton and soliton-antisoliton solutions have been
obtained for the planar integrable chiral model (\ref{1}).
These structures travel with non-constant velocity; their size is
non-constant;  and they interact non-trivially.
Such results might be useful for connecting integrable and
nonintegrable systems which possess soliton solutions.
In addition, they indicate the likely occurrence of new phenomena in
higher dimensional soliton theory that are not present in (1+1) dimensions.

It seems likely that there are many more interesting solutions still to be
found; an open question being what is the general form of the function
$\psi$ when it has a higher-order pole in $\lm$.
One could, for example, investigate the case $n=3$ for $\psi(\lm)$ with a
single and a double pole; and determine  the scattering properties of
the emerging structures, in terms of  their initial velocity
and of the values of the impact parameter.
Finally, it would be of great interest to deduce the general form of the
function $\psi(\lm)$ for the soliton-antisoliton case (\ref{24})
with the only constraint to satisfy the reality condition (\ref{5}) and
the requirement that the matrices $A=(L \psi)\psi^{-1}$ and $B=(M
\psi)\psi^{-1}$ be independent of $\lm$.

\noindent
{\bf Acknowledgments.}

Many thanks are due to R. S. Ward for stimulating suggestions and
discussions and to W. J. Zakrzewski for helpful comments.
I also thank the Department of Mathematical Sciences, of Durham University
for financial support.

\vfill\eject

\newpage

{\bf Figure Captions.}

{\bf Figure 1:} The energy density ${\cal E}$ (16) at increasing times.

{\bf Figure 2:} The energy density ${\cal E}$ (18) at various times.

{\bf Figure 3:} Energy density at various values of $\ve$ for a system of
two solitons ($t=-15$).

{\bf Figure 4:} Energy density at increasing times for a system of
three solitons with $60^0$ angle scattering.

{\bf Figure 5:} Energy density at various times for the scattering of
three solitons, with one being static at the origin.

{\bf Figure 6:} Energy density at increasing times when $\psi(\lm)$ has a
triple pole (and no others).

{\bf Figure 7:} Energy density at increasing times showing a $90^0$
scattering between a soliton and an antisoliton.

{\bf Figure 8:} Topological charge density at increasing times for
(a) soliton-antisoliton scattering, (b) one-soliton  two-antisoliton
scattering.

{\bf Figure 9:} Energy density of a system consisting of a
soliton and  two antisolitons at various times.

\end{document}